\documentclass[12pt]{article}
\usepackage{authblk}
\usepackage{graphicx}
\usepackage{cite}
\usepackage{amssymb,amsmath}
\begin{document}
\title{Molecular dynamics study of the competitive binding of hydrogen
peroxide and water molecules with the DNA phosphate groups}%
\author[1]{S.M. Perepelytsya}
\author[2]{J.~Uli$\check{\mathrm{c}}$n$\acute{\mathrm{y}}$}
\author[1]{S.N.~Volkov}

\affil[1]{Bogolyubov Institute for Theoretical Physics of the National Academy of Sciences of Ukraine, 14b, Metrolohichna Str., Kyiv 03143, Ukraine, perepelytsya@bitp.kiev.ua}
\affil[2]{University of P. J. $\check{\mathrm{S}}$af$\acute{\mathrm{a}}$rik in Ko$\check{\mathrm{s}}$ice, Institute of Physics at Faculty of Science $\acute{\mathrm{U}}$FV PF UPJ$\check{\mathrm{S}}$\\
Park Angelinum 9, 041 54 Ko$\check{\mathrm{s}}$ice, Slovakia}
\setcounter{page}{1}%
\maketitle

\begin{abstract}
The hydrogen peroxide is present in the living cell at small concentrations that increase under the action of the heavy ion beams in the process of anticancer therapy. The interactions of hydrogen peroxide with DNA, proteins and other biological molecules are poorly understood. In the present work the competitive binding of the hydrogen peroxide and water molecules with the DNA double helix backbone has been studied using the molecular dynamics method. The simulations have been carried out for the DNA double helix in a water solution with hydrogen peroxide molecules and Na$^{+}$ counterions. The obtained radial distribution functions of counterions, H$_2$O$_2$ and H$_2$O molecules with respect to the oxygen atoms of DNA phosphate groups have been used for the analysis of the formation of different complexes. The calculated mean residence times show that a hydrogen peroxide molecule stays at least twice as long near the phosphate group (up to 7 ps) than a water molecule (about 3 ps). The hydrogen peroxide molecules form more stable complexes with the phosphate groups of the DNA backbone than water molecules do.
\end{abstract}

\section{Introduction}
Hydrogen peroxide is always present in the living cell at micromolar concentration which is controlled by the specific enzymes \cite{Watson,Halliwell,Malinouski,Tewari,Pedziwiatr}. Due to water radiolysis induced by the action of the heavy ion beams in anticancer therapy the concentration of H$_2$O$_2$ drastically increases in different compartments of a cell, in particular, in the cell nuclei, where the DNA macromolecule is localized \cite{Kraft,Kreipl,Durante2018}. However, the detailed analysis of the action of heavy ion beams on DNA double helix show that the formation of single- and double-strand breaks are not enough to stop the DNA biological functioning, and other mechanisms of the action were proposed \cite{Sol09,Shok1,Shok2,Sol13}. In particular, the blocking of the biologically  active centers of the DNA double helix by hydrogen peroxide molecules examined as an addition mechanism in anticancer therapy \cite{PZPV2015,PZPV2016}. In this regard the interaction of hydrogen peroxide with the structure elements of the DNA double helix is expected to be studied. In the same time, the experimental studies of H$_2$O$_2$-DNA interactions are complicated because H$_2$O$_2$ molecule is thermodynamically unstable under the influence of physical factors and during the measurement process it may decay into radicals. Therefore, the methods of molecular modeling and simulation can give important insights to the mechanisms of interaction of hydrogen peroxide molecule with the DNA double helix.

The DNA macromolecule consists of the nucleotide bases inside and the sugar-phosphate backbone outside the double helix having the minor and major grooves due to the special  twisting of opposite strands around each other \cite{Saenger}. In this regard, hydrogen peroxide can bind to DNA specifically through the atoms of the nucleotide bases in the grooves and non-specifically through the oxygen atoms of the phosphate groups of the macromolecule backbone. The molecular modeling with the use of atom-atom potential functions \cite{PZPV2015,PZPV2016}  and the methods of quantum mechanics of molecules \cite{PiaV2018} show that the complexes of the phosphate groups with hydrogen peroxide are more stable than with water molecules. The possibility of blocking  the DNA specific recognition sites and the opening of the base pairs by the hydrogen peroxide molecules \cite{ZPV2019, ZV2019} were also been pointed out. To perform more realistic modeling of interaction processes   of hydrogen peroxide with the structure elements of the DNA double helix the interactions with the surrounding water molecules and ions should be taken into consideration. At the present time, the atomistic  molecular dynamics is the one method that may provide the detail information and enough accuracy of the research.

To perform the molecular dynamics simulations at the atomistic level the potential functions, describing the intramolecular and intermolecular interactions for all atoms of the system, should be determined. In the standard parameter sets for the potential functions, known as the force fields \cite{Phillips,Foloppe,MacKerell}, the parameters for the hydrogen peroxide molecule are not implemented yet. Therefore, in the molecular dynamics simulations of the systems with H$_2$O$_2$ molecules the parameters were selected in different way \cite{Praprotnik,Campo,Martins,Soloviev}. In the work \cite{Praprotnik} the parameters for intramolecular potentials of the H$_2$O$_2$ molecule were obtained from the comparison of the solutions of the vibrational problem with the experimental spectra of hydrogen peroxide \cite{Giguere}. In the molecular dynamics simulations \cite{Soloviev} for DNA double helix in water solution with hydrogen peroxide molecules the intermolecular parameters for H$_2$O$_2$ molecule were used the same as in TIP3P water model \cite{TIP3P}, and the intramolecular parameters were taken two times lower as the parameters of the work \cite{Praprotnik}. The work \cite{Soloviev}  was focused on a problem of the description of irradiation-induced chemical bond breaks in the atomistic molecular dynamics simulations by the method developed in \cite{Volkov}, therefore the interaction of hydrogen peroxide with the DNA double helix was out the scope of interest. The consistent development of the model of hydrogen peroxide peroxide molecule for the atomistic simulations was performed only recently in \cite{English}, where the parameters for the both intermolecular and intramolecular potentials  are carefully elaborated. Thus, at the present time the study of interaction of hydrogen peroxide molecules with the DNA double helix may be carried out with the use of standard method of atomistic molecular dynamics simulations.

The goal of the present work is to perform the molecular dynamics study of the competitive binding of hydrogen peroxide and water molecules with the phosphate groups of DNA backbone. To solve the problem the molecular dynamics simulations of DNA in a water solution  with H$_2$O$_2$ molecules and sodium ions have been carried out. The simulation details and the methods of the analysis are described in the following section. In the results section the structure of the solvation shell of the phosphate group in complex with the hydrogen peroxide and water molecules are described using the radial distribution functions. In the Discussion section the stability of the complex of the phosphate group with the hydrogen peroxide and water molecules are compared using the derived values of mean residence times. The results show that the interaction of hydrogen peroxide with the oxygen atoms of DNA phosphate group are more favorable than with water molecules.

\section{Materials and methods}
The molecular dynamics simulations have been carried out for the system bearing the infinite \emph{B}-DNA double helix in water solution with hydrogen peroxide and Na$^{+}$ counterions.  The DNA duplex consisted of repeating 20 base pair polynucleotide d(\underline{\emph{CGCGAATT}$\overline{CGCG}$}$\overline{AATTCGCG}$) that has two overlapping motifs of nucleotide sequence (underlined and overlined), each of which is the Drew-Dickerson dodecamer \cite{Drew}. The Drew-Dickerson dodecamer is a model polynucleotide that is often used in the molecular dynamics simulations \cite{Orozco,Perepelytsya2018,SJAF}. It represents different features of \emph{B}-DNA double helix structure, in particular, the dependence of the width of the minor and major grooves on the sequence of nucleotides, formation of the ordered system of water molecules in the minor groove of the double helix (hydration spine) and others. To mimic an infinite double helix the ends of the DNA fragment have been linked with their images in the adjacent boxes.

The DNA double helix has been immersed in a water box with 3711 water molecules and 432 hydrogen peroxide molecules. The number of hydrogen peroxide molecules corresponded to the concentration about ca 17 mass percent. The system has been neutralized using 40 monovalent Na$^{+}$ ions in the simulation box that corresponds to the concentration 416 mM. The initial positions of the counterions were randomly generated to be not within 5 {\AA} between counterions and not within 7 {\AA} to the DNA double helix. The initial positions of hydrogen peroxide molecules were generated to create a distribution of the not overlapped molecules in the simulation box and localized not closer than 2 {\AA} to the atoms of DNA macromolecule. The initial state of the simulated system is shown on the Figure 1.

\begin{figure}
\begin{center}
\resizebox{0.7\textwidth}{!}{%
  \includegraphics{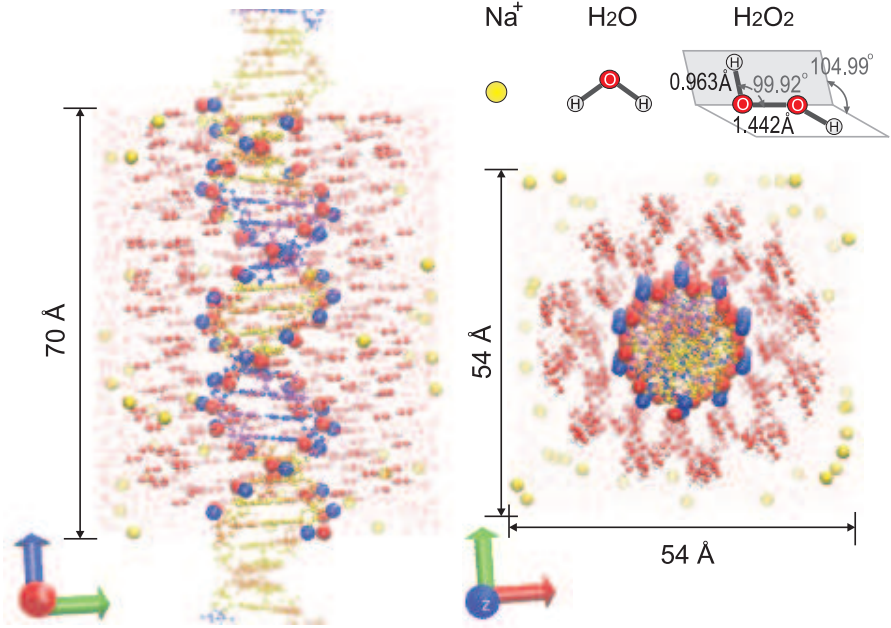}
}
\caption{The infinite DNA double helix in the water box with hydrogen peroxide and Na$^{+}$ counterions. The repetitive fragment of DNA duplex d(CGCGAATTCGCGAATTCGCG): Cytosine (orange), Guanine (yellow), Adenine (blue), Thymine (purple). Oxygen atoms O$_1$ and O$_2$ of the phosphate group of the DNA double helix are shown as enlarged blue and red spheres, respectively. (\emph{Colour online.})}
\end{center}
\end{figure}

In the present work the CHARMM force field was used \cite{Foloppe,MacKerell}. In this force field the potential energy is described as a sum pair interactions that may be written in the following form:
\begin{align}\label{Eq1}
U=\sum_{bonds}{K_{b}(b - b_{0})^2}+\sum_{angles}{K_{\theta}(\theta - \theta_{0})^2}\\\notag
+\sum_{dihedrals}\sum_{n}{{K_{\chi}^{n}\left(1+\cos(n{\chi} - \delta)\right)}}\\\notag
+\sum_{UB}{K_{UB}(S - S_{0})^2}+\sum_{impropers}{K_{\psi}(\psi - \psi_{0})^2}\\\notag
+\sum_{LJ}{\epsilon_{ij}\left[\left(\frac{R_{ij}^{min}}{r_{ij}}\right)^{12}-2\left(\frac{R_{ij}^{min}}{r_{ij}}\right)^{6}\right]}\\\notag
+\sum_{Coul}{K_{C}}{\frac{q_{i}q_{j}}{{\varepsilon}r_{ij}}}.\notag
\end{align}
Here the first term describes the energy of the atoms linked by chemical bond with the length $b$ and the force constant $K_{b}$, $b-b_{0}$ is the change of bond length. The second term describes the energy of deformation of the valence angle $\theta$ with the force constant $K_{\theta}$, $\theta-\theta_0$ is the change of valence angle. The third term is the energy of dihedral angles $\chi$ formed by four sequentially bonded atoms in the molecule, $n$ is the integer number that denotes the periodicity of the rotational barrier (multiplicity), $K_{\chi}^{n}$ is the force constant, $\delta_i$ is the phase angle of the dihedral. The fourth sum in the Equation \ref{Eq1} (Urey-Bradley term) describes the energy of the atoms separated by two covalent bonds at the distance $S$ with the force constant $K_{UB}$, $S-S_{0}$ is the change of the distance between atoms. The fifth term in the formula (\ref{Eq1}) describes bending of some molecular structural elements (improper torsion); here $\psi$ is the angle angle, $K_{\psi}$ is the force constant, and $\psi-\psi_{0}$ is the change of the angle. The last two terms in the Equation (\ref{Eq1}) describe the van der Waals and Coulomb nonbonded interactions. The van der Waals interactions are described by the Lennard-Jones potential with the parameters $\epsilon_{ij}$ and $R_{ij}^{min}$, $r_{ij}$ is the distance between atoms. The Coulomb term is determined by the partial charges  $q_i$ and $q_j$ on the atoms $i$ and $j$, $\varepsilon$ is the dielectric constant and $K_C$ is the Coulomb's constant.

For the nucleic acids the CHARMM36 parameter set was used \cite{Foloppe,MacKerell}. To link  the ends of the DNA fragment with their images in the adjacent boxes the patch LKNA from CHARMM33 parameter set was taken. The TIP3P water model \cite{TIP3P} and Beglov and Roux parameters for Na$^{+}$ \cite{Beglov} ions were used. The parameters, necessary for the simulation of the hydrogen peroxide molecules, are not implemented in the standard force fields. In the same time, the model of hydrogen peroxide molecule for the atomistic molecular dynamics simulations was elaborated in \cite{English}. The parameters for the hydrogen peroxide molecule \cite{English} that are used in the in present work are shown in the Table 1.

\begin{table}
\begin{center}
\noindent\caption{ The parameters for H$_2$O$_2$ molecule \cite{English}.}\vskip3mm\tabcolsep4.5pt
\noindent{\footnotesize
\begin{tabular}{lcccc}
 \hline%
Bonds&$K_b$ (kcal/mole/{\AA}$^{2}$)&$b_{0}$ (\AA)&&\\%
O--O&285.500&1.442&&\\%
O--H&521.000&0.963&&\\[3mm]
Angles&$K_{\theta}$ (kcal/mole/rad$^2$)&$\theta_{0}$ ($^{\circ}$)&&\\%
H--O--O&60.400&99.92&&\\[3mm]
Dihedral&$K_{\chi}$ (kcal/mole)&$n$&$\delta$ ($^{\circ}$)&\\%
H--O--O--H&2.02&2&0.00&\\[3mm]
Long-range&$\epsilon_{j}$ (kcal/mole)&$R^{min}_{j}/2$ (\AA)&$q$ ($e$)&\\%
O&-0.20384&1.67423&-0.41&\\%
H&-0.046000&0.224500&0.41&\\\hline
\end{tabular}
}
\end{center}
\end{table}

The computer simulations have been performed using the NAMD software package \cite{Phillips}. The lengths of all bonds with hydrogen atoms were constrained using the SHAKE algorithm \cite{SHAKE}. The long-range electrostatic interactions were treated using particle mesh Ewald method \cite{PME}. The switching and cut-off distances for the long-range interactions were set to 8 {\AA} and 10 {\AA}, respectively. The integration time step was 2 fs.  The temperature was kept at 300 $^\circ$K using the Langevin thermostat for all heavy atoms (damping constant 5 ps$^{-1}$). The oscillation time and damping time constants for the Langevin piston are 100 fs and 50 fs, respectively.

The simulation protocol was taken similar to \cite{SJAF}. The simulations were started with the minimization of the system with fixed heavy atoms (only hydrogen atoms are allowed to move). Then the system was minimized with fixed DNA, hydrogen peroxide atoms, and ions. After minimization the water component of the solvent was heated to the temperature 300 $^{\circ}$K and equilibrated  during 25 ps. The time step was 0.5 fs. Then the system with fixed DNA and hydrogen peroxide atoms, and constrained ions was heated to the temperature 300 $^{\circ}$K and then equilibrated with constrained ions (25 ps) and free ions (25 ps). After that the system was minimized with fixed DNA atoms and then  it was heated to 300 $^{\circ}$K with following equilibration with constrained (250 ps) and free (250 ps) hydrogen peroxide atoms. When preparing the solution, the defreezing of the DNA atoms was started: the system was minimized and after that heated to 300 $^{\circ}$K with following equilibration with constrained  DNA atoms (250 ps), than with constrained ends of the DNA fragment (250 ps), and then with all atoms without any constrains (3,75 ns). The simulations up to this stage were performed as NPT ensemble with the constant pressure 101325 Pa. At the production stage the system was simulated at constant volume and temperature (NVT ensemble) with the time step 2 fs during 200 ns. In the present work the first 100 ns interval of the simulation trajectory has been neglected as a time period necessary for  equilibration of the system. Such trajectory interval, obtained with the above simulation protocol, has been shown to be sufficiently enough for the equilibration of all components of the system \cite{Perepelytsya2018}. The analysis of obtained simulation trajectories has been carried out using VMD software package \cite{Humphrey}.

\section{Results}
The structures of the complexes of counterions, water molecules, and hydrogen peroxide molecules with the phosphate groups of the DNA backbone were studied using the radial distribution functions (RDFs). The RDFs were built with respect to the oxygen atoms O$_1$ and O$_2$ of the phosphate groups (Fig. 2). These atoms are not equivalent in the structure of the DNA double helix, and being exposed into the solution they are accessible for the interaction with the molecules and ions. Thus, the radial distribution functions for oxygen atoms of the hydrogen peroxide molecules (RDF$_{PER}^{O1}$ and RDF$_{PER}^{O2}$), water molecules (RDF$_{W}^{O1}$ and RDF$_{W}^{O2}$), and Na$^+$ counterions (RDF$_{Ion}^{O1}$ and RDF$_{Ion}^{O2}$) were calculated. The RDFs were calculated using the plug-in \cite{gr} implemented to VMD \cite{Humphrey}.

\begin{figure}
\begin{center}
\resizebox{0.7\textwidth}{!}{%
  \includegraphics{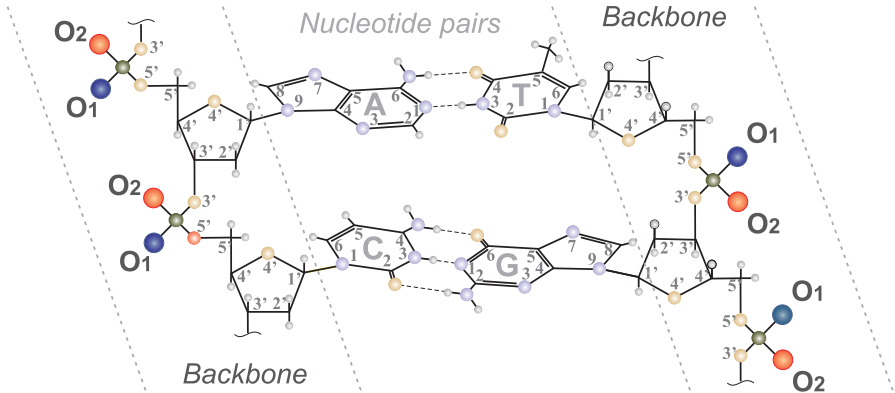}
}
\caption{The nucleotide pairs. Oxygen atoms O$_1$ and O$_2$ of the phosphate group of the DNA backbone are shown as enlarged blue and red spheres, respectively. (\emph{Colour online.})}
\end{center}
\end{figure}

The radial distribution functions, characterizing the position of hydrogen peroxide molecule with respect to the phosphate group of the backbone of DNA double helix, are shown on the Figure 3. The RDF$_{PER}^{O1}$ and RDF$_{PER}^{O2}$ have two peaks, the first is at about 2.6 {\AA} and the second one at about 3.3 {\AA}. The first peak is about twice higher than the second one.  To determine the average number of hydrogen peroxide molecules around the O$_1$ or O$_2$ atoms of DNA phosphate groups the  integrals below the RDFs were calculated. The integral below the first and the second RDFs peaks are known as the first and the second coordination numbers, $n_1$ and $n_2$, respectively. The dependencies of the integrals  of RDFs  show that the first and the second coordination numbers have about the same values $n_1{\approx}n_1$ (Fig. 3). Ergo, the number of oxygen atoms of hydrogen peroxide molecule that are at the distances of the first and the second peaks of the radial distribution function is the same. This is possible if the both peaks of the radial distribution function are originated from the oxygen atoms that belong to the same H$_2$O$_2$  molecule and not to different ones. The coordination numbers in the case of O$_2$ atom of the phosphate group are about 50 percent higher than in the case of O$_1$ atom, indicating that the hydrogen peroxide molecule interacts more intensively with O$_2$ atom of PO$_4^{-}$ group.

\begin{figure}
\begin{center}
\resizebox{0.55\textwidth}{!}{%
  \includegraphics{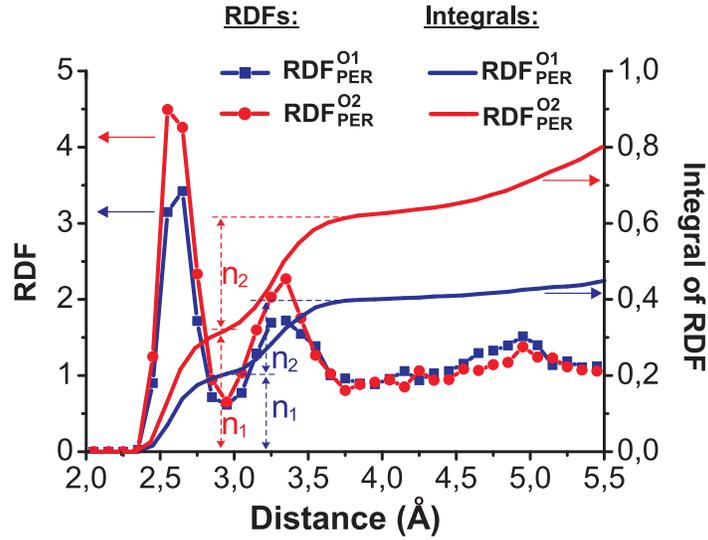}
}
\caption{Radial distribution functions and their integrals for the oxygen atoms of hydrogen peroxide molecules with respect to O$_1$ and O$_2$ atoms of the phosphate groups of the DNA backbone, RDF$_{PER}^{O1}$ and RDF$_{PER}^{O2}$, respectively. The dashed arrows indicate the first and the second coordination numbers $n_1$ and  $n_2$, respectively. (\emph{Colour online.})}
\end{center}
\end{figure}

To characterize distribution of water molecules around DNA phosphate groups the radial distribution functions of oxygen atoms of water molecules with respect to O$_1$ and O$_2$ atoms of the phosphate group have been built (Fig. 4). The resulted RDF$_{W}^{O1}$ and RDF$_{W}^{O2}$ are characterized by the intensive peek at 2.7 \AA, related to the first hydration shell of the phosphate group, and a broad  band from 3.5 to 5.5 \AA, related to the second hydration shell. The integrals of RDFs show that the first coordination number of water molecules near one oxygen atoms of the phosphate group are within $n_1=(2\div2.3)$. The atom O$_1$  has slightly higher coordination number than the atom O$_2$.

\begin{figure}
\begin{center}
\resizebox{0.55\textwidth}{!}{%
  \includegraphics{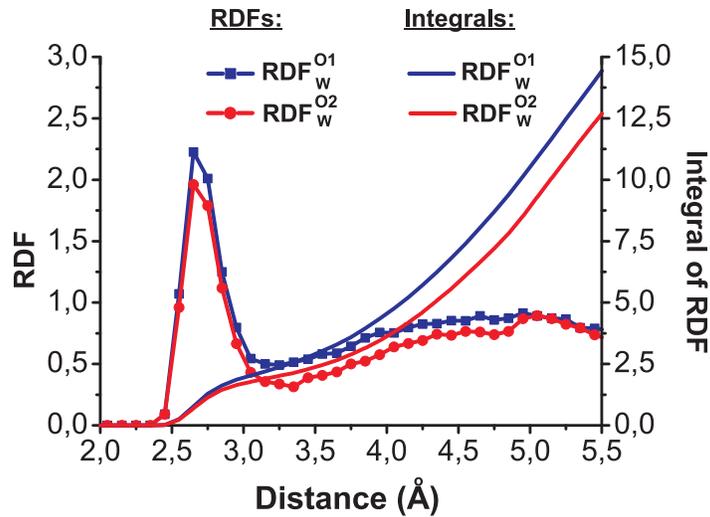}
}
\caption{Radial distribution functions RDF$_{W}^{O_1}$ and RDF$_{W}^{O_2}$ and their integrals for the oxygen atoms of water molecules with respect to the O$_1$ and O$_2$ atoms of the phosphate groups of the DNA backbone, respectively. (\emph{Colour online.})}
\end{center}
\end{figure}

To characterize the distribution of counterions around the phosphate groups of DNA double helix the radial distribution functions of Na$^{+}$ with respect to O$_1$ and O$_2$ atoms of the phosphate group have been built (Fig. 5). The obtained RDF$_{Ion}^{O1}$ and RDF$_{Ion}^{O2}$ are characterized by two peeks at the distance 2.25 {\AA} and about 4.6 {\AA}. The first intense peak is related to the counterions that form direct contacts with oxygen atoms of the phosphate group and in the case of the O$_1$ atom it is much more intense than in the case of the O$_2$ atom. The second small peak is related to the counterions that form water-mediated contacts with the oxygen atoms of the phosphate groups and in the case of the O$_1$ atom it is less intense than in the case of the O$_2$ atom. The integrals for the first peak of the RDFs show are about 0.05 and 0.01 for O$_1$ and O$_2$ atoms of the phosphate group, respectively. In the same time integrals for the second peak of the RDFs have the values 0.15 and 0.3 for O$_1$ and O$_2$ atoms of the phosphate group, respectively. The calculated coordination numbers indicate that the O$_1$ atom of the phosphate group is more accessible for direct contact with  Na$^{+}$ ions. Note,  at the distance 7.5 {\AA}  the integral of RDFs results the coordination number about 0.8 that is close to the value predicted by the counterion condensation theory \cite{Manning,FK}.

\begin{figure}
\begin{center}
\resizebox{0.55\textwidth}{!}{%
  \includegraphics{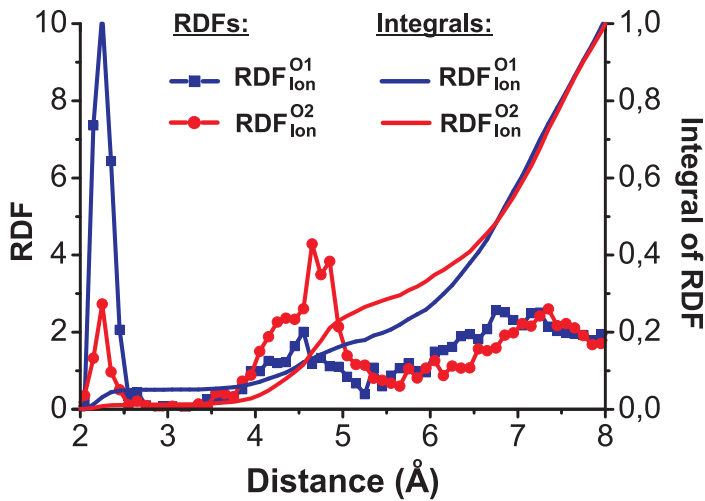}
}
\caption{Radial distribution functions and their integrals for Na$^{+}$ counterions with respect to the O$_1$ and O$_2$ atoms of the phosphate groups of the DNA backbone, RDF$_{Ion}^{O_1}$ and RDF$_{Ion}^{O_2}$, respectively. (\emph{Colour online.})}
\end{center}
\end{figure}

\section{Discussion}
The obtained radial distribution functions show that the structure of the solvation shell of the phosphate groups of the DNA in water solution with hydrogen peroxide molecules and Na$^+$ counterions is characterized by nonuniform interaction with the atoms of O$_1$ and O$_2$ that is the the result of the double helix structure. The hydrogen peroxide molecules compete with water molecules and counterions for the interaction sites that are O$_1$ and O$_2$ atoms of the phosphate group. In the same time, the obtained RDFs show that there is a correlation: hydrogen peroxide molecules prefer to bind with the  O$_2$ atoms, while water molecules bind to  O$_1$ more extensively. In this regard, it is important to understand which of the complexes, with the participation of H$_2$O$_2$ or H$_2$O, is more stable.

To characterize the stability of the complexes of H$_2$O$_2$ and H$_2$O molecule with the oxygen atoms of the phosphate group the the potentials of mean force (PMF) have been calculated from the radial distribution functions as follows:
\begin{equation}\label{Eq2}
U(r)=-k_{B}T\ln{g(r)},
\end{equation}
where $k_{B}$ is the Boltzmann constant, $T$ is the temperature. The obtained PMF for hydrogen peroxide molecule, water molecules, and counterions near the phosphate group of the DNA backbone are shown on the Figure 6.

\begin{figure}[h]
\begin{center}
\resizebox{1\textwidth}{!}{%
  \includegraphics{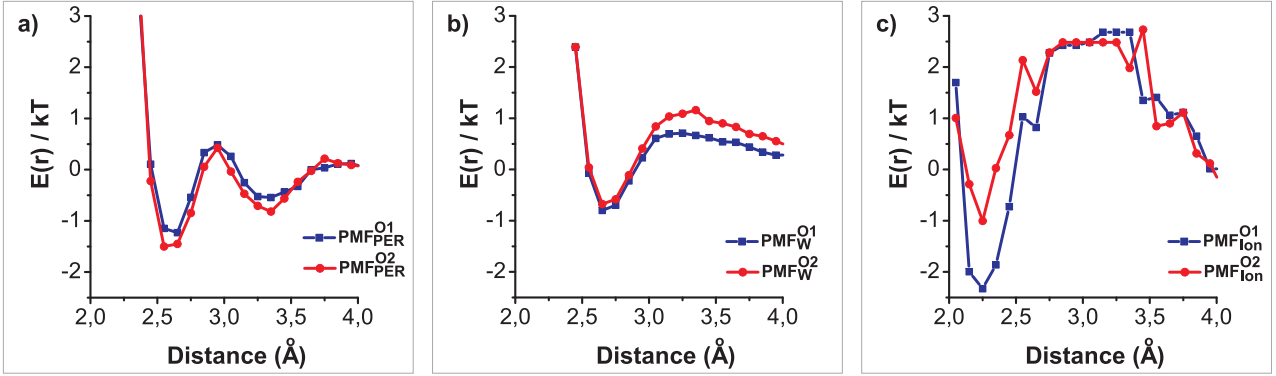}
}
\caption{Potentials of mean force for the oxygen atoms of water molecules (PMF$_{W}^{O1}$ and PMF$_{W}^{O2}$) (a), hydrogen peroxide molecules (PMF$_{PER}^{O1}$ and PMF$_{PER}^{O2}$) (b), and Na$^{+}$ counterions (PMF$_{Ion}^{O1}$ and PMF$_{Ion}^{O2}$) (c) with respect to O$_1$ and O$_2$ atoms of the phosphate group. (\emph{Colour online.})}
\end{center}
\end{figure}

In the case of hydrogen peroxide there are two potential wells, the first is characterized by the potential well with the potential barrier more than 1 kcal$/$mole, and the second one is about 0.3 kcal$/$mole. In the case of water molecules the obtained PMF are characterized by one deep potential well with the potential barrier about 1 kcal$/$mole. The PMF for counterions is characterized by the deepest potential well (more than 2.5 kcal$/$mole) that is due to the electrostatic attraction between negatively charged phosphate group and positively charged Na$^{+}$ ion. The values of the potential barriers of obtained are shown in the Table 1.

\begin{table}
\begin{center}
\noindent\caption{The values of the potential barriers ${\triangle}E$ and the residence times $\tau$ for water and hydrogen peroxide molecules, and Na$^{+}$ counterions near O$_1$ and O$_2$ atoms of the phosphate group.}\vskip3mm\tabcolsep4.5pt
\noindent{\footnotesize
\begin{tabular}{lcc}\\[2mm]%
\hline%
&  atom O$_1$  & atom O$_2$  \\[2mm]%
& \multicolumn{1}{c}{${\triangle}E$ (kcal$/$mole)} {$\tau$ (ps)} & \multicolumn{1}{c}{${\triangle}E$ (kcal$/$mole)}{$\tau$ (ps)}\\[2mm]%
\hline%
  H$_2$O$_2$   & \multicolumn{1}{c}{1.49} {4.83}  & \multicolumn{1}{c}{1.68} {6.46} \\
  H$_2$O       & \multicolumn{1}{c}{0.94} {1.91}  & \multicolumn{1}{c}{1.13} {2.66} \\
  Na$^{+}$     & \multicolumn{1}{c}{3.16} {55.91} & \multicolumn{1}{c}{1.95} {9.44}\\[2mm]%
\hline
\end{tabular}
}
\end{center}
\end{table}

Using the determined values of the potential barriers $\Delta{E}$ the average residence time $\tau$ for water and hydrogen peroxide molecules may be estimated using the equation of Arheniuns type \cite{Ismailov}:
\begin{equation}\label{Eq3}
\tau=2\tau_{0}\exp{\left(\frac{\Delta{E}}{k_{B}T}\right)},
\end{equation}
where $\tau_{0}$ is the characteristic time of approaching of the molecule to the potential barrier $\Delta{E}$. The factor 2 in the formula (\ref{Eq3}) appears because in our approach we consider that  being at the top of the potential barrier the water molecule may leave the hydration shell or return back to the ion with the equal probability. The value of $\tau_{0}$ may be estimated as the period of vibrations: $\tau_{0}=2\pi\sqrt{m/k}$, where $m$ is the mass of a molecule or the ion, and $k$ is the force constant that may be determined from PMF fitting.

The obtained values of the mean residence times show that the hydrogen peroxide molecules reside near the oxygen atoms of the phosphate groups during the time up to 7 ps, while the water molecules have more than twice lower $\tau$ values (Table 2). The largest values of the residence times are in the case of Na$^+$ counterions (up to 60 ps). The values of the residence times are different in the case of O$_1$ and O$_2$ atoms of the phosphate group, and H$_2$O$_2$ and H$_2$O molecules stay longer near the atom O$_1$, while Na$^+$ counterions near the atom O$_2$.  Thus the complex of the phosphate group of DNA double helix with hydrogen peroxide molecule has more than twice longer lifetime than with water molecule which means that H$_2$O$_2$ is more tightly bonded to the  PO$_{4}^{-}$ group than H$_2$O. This result confirms the prediction, made in the previous calculations \cite{PZPV2015,PZPV2016,PiaV2018}, and supports the idea about the possible influence of hydrogen peroxide molecule on the mechanisms of the DNA biological functioning in the living cell.

\section{Conclusions}
The molecular dynamics simulation of the DNA macromolecule in a water solution with hydrogen peroxide molecules and Na$^{+}$ counterions have been carried out to characterize the competitive binding of hydrogen peroxide and water molecules with the phosphate groups of the double helix backbone. The results show that the oxygen atoms O$_1$ and O$_2$ of DNA phosphate groups are not equivalent with respect to the interaction with the solution components. The water molecules interact more intensively with O$_1$ atom, while the hydrogen peroxide molecules  with O$_2$  atom. The counterions bind mostly to the O$_2$ atom. The complexes of H$_2$O$_2$ molecules with the phosphate groups are characterised by the lifetime up to 7 ps that is more than twice longer than in the case of the complexes with H$_2$O molecules (about 2.5 ps). Thus, the hydrogen peroxide molecules form more stable complexes with the phosphate groups of the DNA backbone than water molecules that is in agreement with previous calculations  \cite{PZPV2015,PZPV2016,PiaV2018} indicating to the possible influence of hydrogen peroxide molecule on the mechanisms of the DNA biological functioning in the living cell.

\vskip3mm \textit{Acknowledgement.}
The present work was partially supported by the Project  of the National Academy of Sciences of Ukraine (0120U100858). The authors gratefully acknowledge the computational facilities of the Department of Biophysics of the University of P. J. \v{S}af\'{a}rik in Ko\v{s}ice (CELIMKBF cluster).

\end{document}